\documentclass[aps,showpacs,showkeys,twocolumn,groupedaddress]{revtex4}
\usepackage{graphicx}
\def\aprle{\buildrel < \over {_{\sim}}}

\begin{document}
\title{\vspace*{-0.5in}
	\hfill {\rm FERMILAB-Pub-01/374-T}\\[0.2in]
	Comparison of LMA and LOW Solar Solution Predictions 
	in an SO(10) GUT Model}

\author{Carl H. Albright}
\email[E-mail address: ]{albright@fnal.gov}
\affiliation{Department of Physics, Northern Illinois University, 
	DeKalb, IL 60115, USA}
\affiliation{Fermi National Accelerator Laboratory, P.O. Box 500,
	Batavia, IL 60510, USA}
\author{S. Geer}
\email[E-mail address: ]{sgeer@fnal.gov}
\affiliation{Fermi National Accelerator Laboratory, P.O. Box 500,
	Batavia, IL 60510, USA}

\date{\today}

\begin{abstract}
Within the framework of an SO(10) GUT model that can accommodate both the 
LMA and LOW solar neutrino mixing solutions by appropriate choice of the
right-handed Majorana matrix elements, we present explicit predictions for the 
neutrino oscillation parameters $\Delta m^2_{21}$, $\sin^2 2\theta_{12}$, 
$\sin^2 2\theta_{23}$, $\sin^2 2\theta_{13}$, and $\delta_{CP}$.  Given the 
observed near maximality of the atmospheric mixing, the model favors the 
LMA solution and predicts that $\delta_{CP}$ is small.  The suitability of 
Neutrino Superbeams and Neutrino Factories for precision tests of the two 
model versions is discussed.
\end{abstract}
\maketitle
\vspace*{-0.2in}
Over the last few years the evidence for neutrino oscillations between
the three known neutrino flavors
($\nu_e, \nu_\mu$, and $\nu_\tau$) has become increasingly convincing. The
atmospheric neutrino flux measurements from
the Super-Kamiokande (Super-K) experiment exhibit a deficit of muon neutrinos
which varies with zenith angle (and hence baseline) in a way consistent with
$\nu_\mu \to \nu_\tau$ oscillations \cite{atm}.  In addition, recent combined 
evidence from Super-K and the SNO experiments \cite{solar} indicate that some 
electron-neutrinos from the sun are oscillating into muon and/or tau neutrinos.
While the atmospheric neutrino data with $\nu_\mu \to \nu_\tau$
oscillations points to a small region of the mixing parameter space \cite{atm},
the solar neutrino data is consistent with at least two regions of parameter 
space \cite{solarregions}, corresponding to either the Large Mixing Angle (LMA) 
or to the LOW  MSW \cite{MSW} solution.  

Neutrino oscillation data constrain Grand Unified Theories (GUTs) which 
provide a theory of flavor and relate lepton masses and mixings to quark 
masses and mixings.  It is known that the presently implied neutrino mass 
scales can be accommodated naturally within the framework of GUTs by the 
seesaw mechanism \cite{gmrsy}.  In practice finding an explicit GUT model 
for the LMA solution has been found challenging.  However, one 
example has been constructed by Barr and one us \cite{ab}.  In this model
the Dirac and Majorana neutrino mass matrices are intimately related.  
It has been shown by the present authors \cite{ag} that by varying the 
Majorana mass matrix parameters any point in the presently-allowed LMA region
can be accommodated.  In the present paper we show 
that the LOW region can also be realized in the model by choosing an 
appropriate texture for the right-handed Majorana mass matrix.  In addition 
to the Majorana mass matrix, we also spell out the Dirac mass matrices, list 
the values of the associated input parameters and give results for the quark 
and charged lepton sectors.  For each choice of the Majorana matrix we discuss
the oscillation predictions.  We use our results to illustrate how Neutrino 
Superbeams and Neutrino Factories \cite{geer} can further test this GUT model.
Our results suggest that, independent of which is the preferred solution 
(LMA or LOW), Neutrino Superbeams and Factories will be necessary to identify 
the correct model, and that further investment in developing them should be 
encouraged.

Within the framework of three-flavor mixing, the flavor eigenstates 
$\nu_\alpha\ (\alpha = e, \mu, \tau)$ are related to
the mass eigenstates $\nu_j\ (j = 1, 2, 3)$ in vacuum by
\begin{equation}
\nu_\alpha = \sum_j U_{\alpha j} \nu_j \;,\quad U \equiv U_{MNS}\Phi_M
\end{equation}
where $U$ is the unitary $3\times3$ Maki-Nakagawa-Sakata (MNS) mixing matrix
\cite{MNS} times a diagonal phase matrix $\Phi_M = {\rm diag}(e^{i\chi_1},
\ e^{i\chi_2},\ 1)$. 
The MNS matrix is conventionally parametrized by 3 mixing angles 
($\theta_{23}, \theta_{12}, \theta_{13}$) and a CP-violating phase,
$\delta_{CP}$:
\begin{equation}
\begin{array}{l}
  U_{MNS} =\\[4pt]
\left(\matrix{c_{12}c_{13} & s_{12}c_{13} & s_{13}\xi^* \cr  
	-s_{12}c_{23} - c_{12}s_{23}s_{13}\xi &
        c_{12}c_{23} - s_{12}s_{23}s_{13}\xi & s_{23}c_{13}\cr
        s_{12}s_{23} - c_{12}c_{23}s_{13}\xi &
        -c_{12}s_{23} - s_{12}c_{23}s_{13}\xi & c_{23}c_{13}\cr}\right)\\
\end{array}
\label{eq:mixing}
\end{equation}
where $c_{jk} \equiv \cos\theta_{jk},\ s_{jk} \equiv \sin\theta_{jk}$
and $\xi = e^{i\delta_{CP}}$.
The three angles can be restricted to the first quadrant,
$0\le \theta_{ij} \le \pi/2$, with $\delta_{CP}$ in the range $-\pi \le 
\delta_{CP} \le \pi$, though it proves advantageous to consider 
$\theta_{13}$ in the fourth quadrant for the LMA solutions.  

The atmospheric neutrino oscillation data indicate that \cite{atm}
\begin{equation}
\begin{array}{rcl}
        \Delta m^2_{32} & \simeq& 3.0 \times 10^{-3}\ {\rm eV^2},\\[4pt]
        \sin^2 2\theta_{atm} & =& 1.0,\ (\geq 0.89 \ {\rm at}\ 90\%\ 
		{\rm c.l.}),\\
\end{array}\label{eq:atm}
\end{equation}
where $\Delta m^2_{ij} \equiv m^2_i - m^2_j$ and $m_1,\ m_2$ and $m_3$ are
the mass eigenstates.
The atmospheric neutrino oscillation amplitude can be expressed solely in 
terms of the $U_{MNS}$ matrix elements and is given by
$\sin^2 2\theta_{atm} = 4|U_{\mu 3}|^2(1 - |U_{\mu 3}|^2) \simeq 
4|U_{\mu 3}|^2 |U_{\tau 3}|^2 = c^4_{13}\sin^2 2\theta_{23}$.  The 
approximation is valid because $|U_{e3}|$ is known to be small \cite{CHOOZ}.

The solar neutrino oscillation data from Super-K indicate that,
for the LMA solution, the allowed region is approximately bounded by 
\begin{equation}
\begin{array}{rcl}
        \Delta m^2_{21} & \simeq& (2.2 - 17) \times 10^{-5}\ {\rm eV^2},\\[4pt]
        \sin^2 2\theta_{sol} & \simeq& (0.6 - 0.9),\\
\end{array}\label{eq:LMA}
\end{equation}
while for the LOW solution, 
\begin{equation}
\begin{array}{rcl}
        \Delta m^2_{21} & \simeq& (0.3 - 2) \times 10^{-7}\ {\rm eV^2},\\[4pt]
        \tan^2 \theta_{12} & \simeq& (0.6 - 1.2),\\
\end{array}\label{eq:LOW}
\end{equation}
where the solar neutrino oscillation amplitude is 
$\sin^2 2\theta_{sol} = 4|U_{e1}|^2 (1 - |U_{e1}|^2) \simeq 
4|U_{e1}|^2 |U_{e2}|^2$, while $\tan^2 \theta_{12} = |U_{e2}/U_{e1}|^2$.  

The GUT model we consider is based on an $SO(10)$ GUT with a $U(1) \times 
Z_2 \times Z_2$ flavor symmetry.  The model 
involves a minimum set of Higgs fields which solves the doublet-triplet 
splitting problem.  The Higgs superpotential exhibits the $U(1) \times Z_2 
\times Z_2$ symmetry which is 
used for the flavor symmetry of the GUT model.  
Details of the model can be found in \cite{ab}.  We simply note that 
the Dirac mass matrices $U,\ D,\ N,\ L$ for the up quarks, 
down quarks, neutrinos and charged leptons, respectively, are found 
for $\tan \beta \simeq 5$ to be
\begin{equation}
\begin{array}{l}
U = \left(\matrix{ \eta & 0 & 0 \cr
  0 & 0 & \epsilon/3 \cr 0 & - \epsilon/3 & 1\cr} \right),\
  D = \left(\matrix{ 0 & \delta & \delta' e^{i\phi}\cr
  \delta & 0 & \sigma + \epsilon/3  \cr
  \delta' e^{i \phi} & - \epsilon/3 & 1\cr} \right) \\[0.3in]
N = \left(\matrix{ \eta & 0 & 0 \cr 0 & 0 & - \epsilon \cr
        0 & \epsilon & 1\cr} \right),\
  L = \left(\matrix{ 0 & \delta & \delta' e^{i \phi} \cr
  \delta & 0 & -\epsilon \cr \delta' e^{i\phi} &
  \sigma + \epsilon & 1\cr} \right),\\
\end{array}\label{eq:Dirac}
\end{equation}
where $U$ and $N$ are scaled by $M_U$, and $D$ and $L$ are scaled by $M_D$.
All nine quark and charged lepton masses, plus the three CKM angles and CP
phase, are well-fitted with the eight input parameters 
\begin{equation}
\begin{array}{rlrl}
        M_U&\simeq 113\ {\rm GeV},&\qquad M_D&\simeq 1\ {\rm GeV},\\
        \sigma&=1.78,&\qquad \epsilon&=0.145,\\
        \delta&=0.0086,&\qquad \delta'&= 0.0079,\\
        \phi&= 126^\circ,&\qquad \eta&= 8 \times 10^{-6},\\
\end{array}\label{eq:parameters}
\end{equation}
defined at the GUT scale to fit the low scale observables after evolution 
downward from $\Lambda_{GUT}$:
\begin{equation}
\begin{array}{rlrl}
           m_t(m_t) &= 165\ {\rm GeV},\quad & m_{\tau} &= 1.777\ {\rm GeV},
		\\[2pt]
           m_u(1\ {\rm GeV}) &= 4.5\ {\rm MeV},\quad & m_\mu &= 105.7\ 
                {\rm MeV},\\[2pt]
           V_{us} &= 0.220, \quad & m_e &= 0.511\ {\rm MeV},\\[2pt]
           V_{cb} &= 0.0395, \quad & \delta_{CP} &= 64^\circ.\\
\end{array}
\label{eq:input}
\end{equation}
These lead to the following predictions:
\begin{equation}
\begin{array}{ll}
           m_b(m_b) = 4.25\ {\rm GeV},\quad & m_c(m_c) = 1.23\ {\rm GeV},\\[2pt]
           m_s(1\ {\rm GeV}) = 148\ {\rm MeV},\ & m_d(1\ {\rm MeV}) 
                = 7.9\ {\rm MeV},\\[2pt]
           |V_{ub}/V_{cb}| = 0.080,\quad & \sin 2\beta = 0.64.\\
\end{array}
\label{eq:output}
\end{equation}
With no extra phases present, the vertex of the
CKM unitary triangle occurs near the center of the presently allowed region
with $\sin 2\beta \simeq 0.64$, comparing favorably with recent results 
\cite{BB}.  The Hermitian matrices $U^\dagger U,\ D^\dagger D$, and 
$N^\dagger N$ are diagonalized with small left-handed
rotations, $U_U,\ U_D,\ U_N$, respectively, while $L^\dagger L$ is diagonalized
by a large left-handed rotation, $U_L$.
This accounts for the small value of $|V_{cb}| = |(U^\dagger_U U_D)_{cb}|$,
while $|U_{\mu 3}| = |(U^\dagger_L U_\nu)_{\mu 3}|$ will turn out to be 
large for any reasonable right-handed Majorana mass matrix, $M_R$ \cite{abb}.

The effective light neutrino mass matrix, $M_\nu$, is obtained from the seesaw 
mechanism once $M_R$ is specified.  
While the large atmospheric neutrino mixing $\nu_\mu \leftrightarrow 
\nu_\tau$ arises primarily from the structure of the charged lepton mass 
matrix, the structure of $M_R$ determines the
type of $\nu_e \leftrightarrow \nu_\mu,\ \nu_\tau$ solar neutrino mixing.

To obtain the LMA solution requires some fine-tuning and a hierarchical 
structure for $M_R$, but this can be explained in terms of Froggatt-Nielsen 
diagrams \cite{fn}.  Here we restrict our attention to a slightly less general
form for $M_R$ than that considered in \cite{ab} and \cite{ag}:
\begin{equation}
          M_R = \left(\matrix{b^2 \eta^2 & -b\epsilon\eta & a\eta\cr
                -b\epsilon\eta & \epsilon^2 & -\epsilon\cr
                a\eta & -\epsilon & 1\cr}\right)\Lambda_R,\\
\label{eq:MRLMA}
\end{equation}
where the parameters $\epsilon$ and $\eta$ are those introduced in 
Eq.(\ref{eq:Dirac}) for the Dirac sector.  This structure for
$M_R$ can be understood as arising from one Higgs singlet which induces a 
$\Delta L = 2$ transition and contributes to all nine matrix elements while, 
by virtue of its flavor charge assignment, a second Higgs singlet breaks 
lepton number but modifies only the 13 and 31 elements of $M_R$.  As shown 
in detail in \cite{ag}, we can introduce additional CP violation by assigning a 
relative phase to the two lepton number breaking Higgs singlets, whereby we
set 
\begin{equation}
	a = b - a'e^{i\phi'}.
\label{eq:acomplex}
\end{equation}

On the other hand, we find the LOW solution can be obtained with the simple 
hierarchical structure for $M_R$,
\begin{equation}
	M_R = \left(\matrix{e & d & 0\cr d & 0 & 0\cr 0 & 0 & 1\cr}\right)
		\Lambda_R,\\
\label{eq:MRLOW}
\end{equation}
where by the flavor charge assignments, one Higgs singlet inducing a
$\Delta L = 2$ transition contributes to the 12, 21 and 33 elements, while
a second Higgs singlet also breaks lepton number but contributes only to the
11 matrix element.  For simplicity we keep both $d$ and $e$ real, since the 
leptonic CP phase is inaccessible to measurement for $\Delta m^2_{21}$ values 
in the LOW region.

\begin{figure*}[t!]
\includegraphics*[width=2.4in]{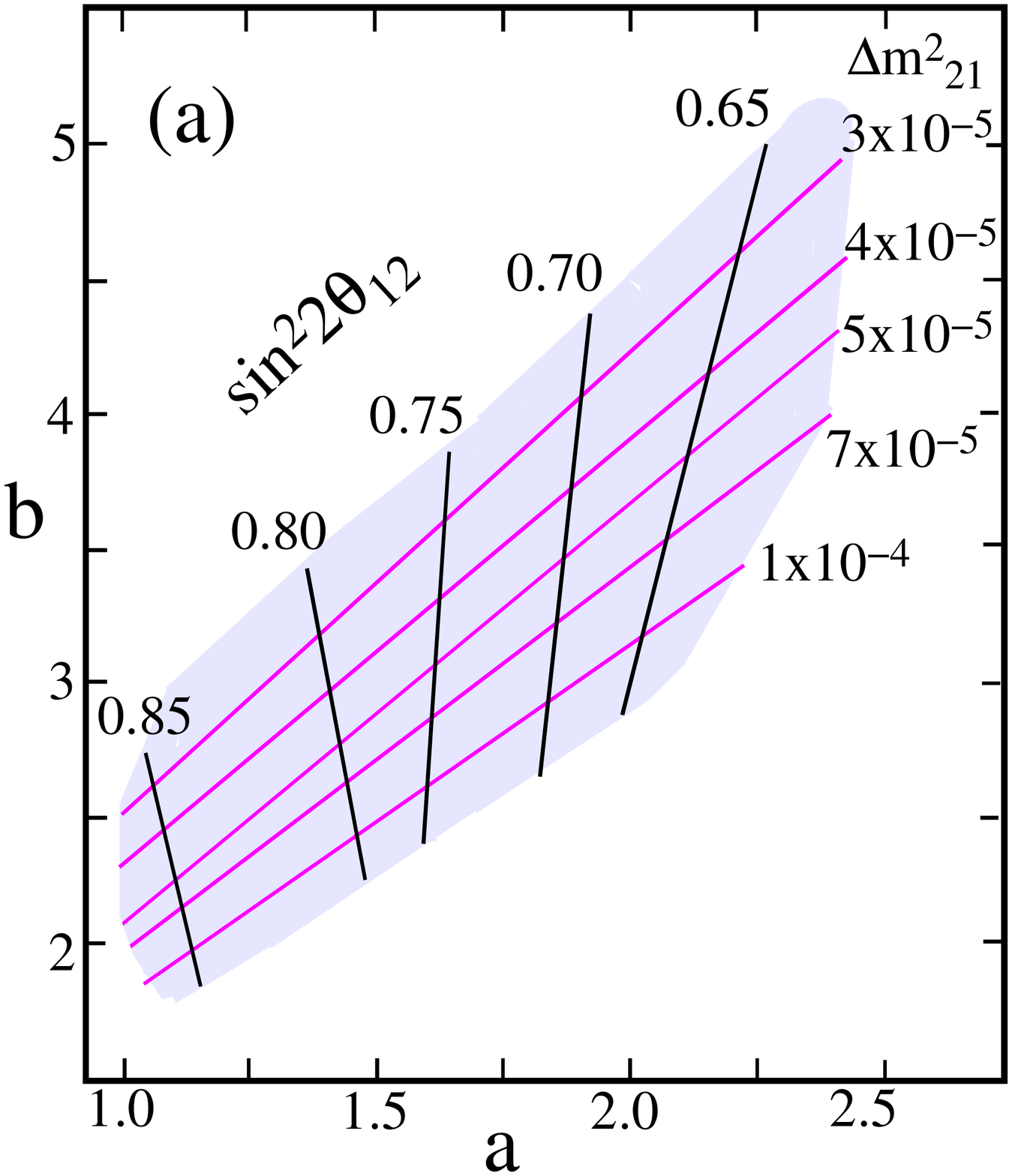}
\hspace*{0.9in}
\includegraphics*[width=2.4in]{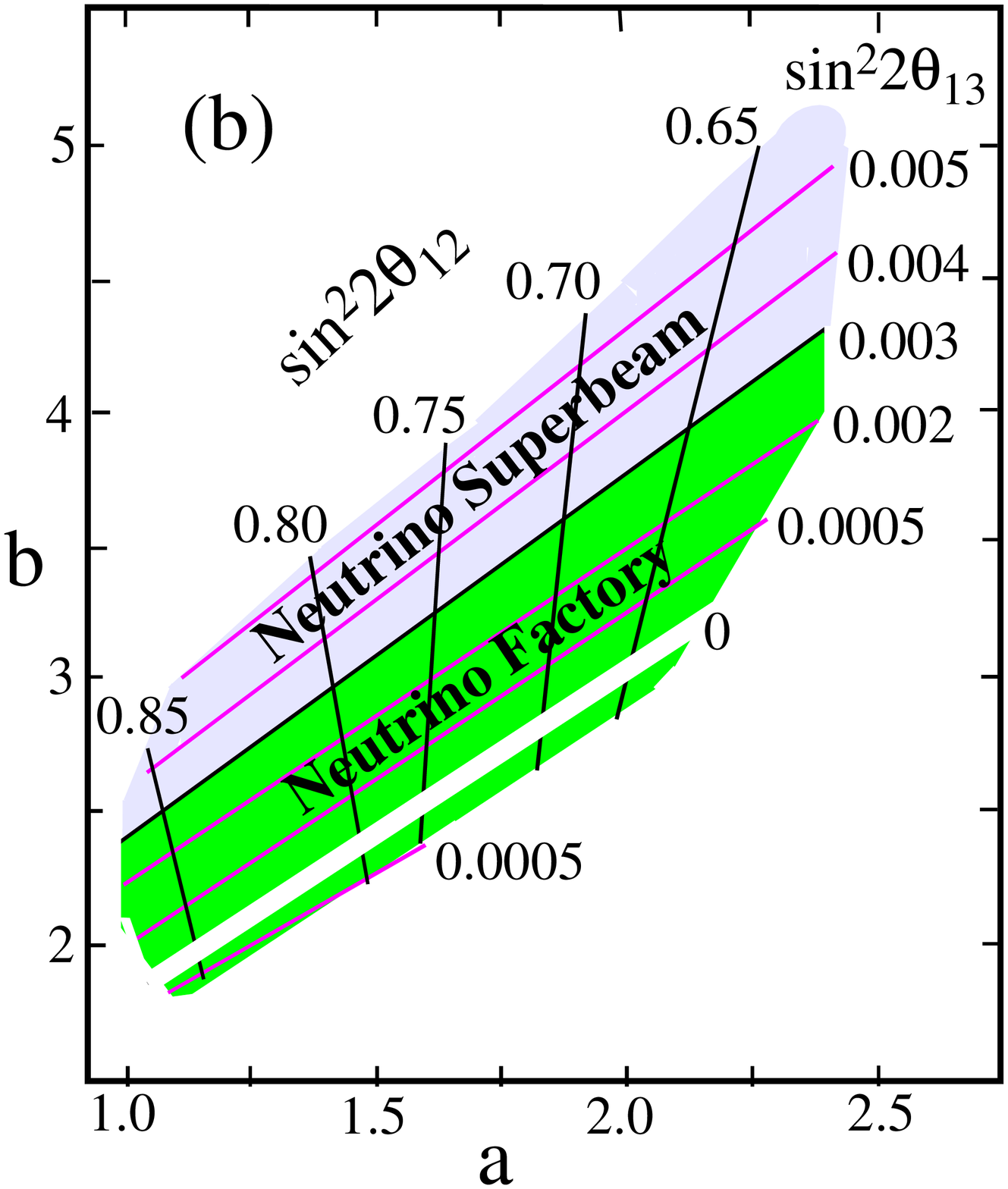}
\vspace*{-0.1in}
\caption[]{The viable region of GUT parameter space consistent 
with the present bounds on the LMA MSW solution.  Contours of 
constant $\sin^2 2\theta_{12}$ are shown together with (a) contours of 
constant $\Delta m^2_{21}$ and (b) contours of $\sin^2 2\theta_{13}$.}
\label{fig:gut_plot}
\end{figure*}
%
\begin{table*}[floatfix]
\caption[]{\label{tab:x1}
List of four points selected in the LMA allowed parameter region to 
illustrate the neutrino oscillation parameter predictions of the GUT model.
Here the CP phase $\delta_{CP}$ arises from $\phi$ in $L$ alone, as no phase
$\phi'$ has been introduced in $M_R$.\\}
\begin{tabular}{ccrcccclr}
  \hspace*{0.2in}$a$ &\hspace*{0.2in} $b$ &\hspace*{0.3in} 
	$\Delta m^2_{21}\ (eV^2)$\hspace*{0.4in} &
        $\Delta m^2_{32}\ (eV^2)$ &\hspace*{0.2in} $\tan^2 \theta_{12}$ 
	&\hspace*{0.2in} $\sin^2 2\theta_{12}$&
        \hspace*{0.2in} $\sin^2 2\theta_{23}$ &\hspace*{0.2in}
        $\sin^2 2\theta_{13}$ &\hspace*{0.2in} $\delta_{CP}$\\ 
	\hline 
  \hspace*{0.15in} 1.0 &\hspace*{0.2in} 2.0 & $6.5 \times 10^{-5}
	$\hspace*{0.45in} & $3.2 \times 10^{-3}$ 
        &\hspace*{0.2in} 0.49 &\hspace*{0.2in} 0.88&\hspace*{0.2in} 0.994
	&\hspace*{0.25in} 0.0008 &\hspace*{0.2in} $-4^\circ$\\
  \hspace*{0.15in} 1.2 &\hspace*{0.2in} 2.8 & $3.3 \times 10^{-5}
	$\hspace*{0.45in} & $3.2 \times 10^{-3}$ 
        &\hspace*{0.2in} 0.43 &\hspace*{0.2in} 0.84&\hspace*{0.2in} 0.980
	&\hspace*{0.25in} 0.0038 &\hspace*{0.2in} $-1^\circ$\\
  \hspace*{0.15in} 1.6 &\hspace*{0.2in} 2.9 & $6.1 \times 10^{-5}
	$\hspace*{0.45in} & $3.2 \times 10^{-3}$ 
        &\hspace*{0.2in} 0.35 &\hspace*{0.2in} 0.77&\hspace*{0.2in} 0.998
	&\hspace*{0.25in} 0.0015 &\hspace*{0.2in} $-3^\circ$\\
  \hspace*{0.15in} 1.7 &\hspace*{0.2in} 2.7 & $10.9 \times 10^{-5}
	$\hspace*{0.45in} & $3.2 \times 10^{-3}$ 
        &\hspace*{0.2in} 0.32 &\hspace*{0.2in} 0.73&\hspace*{0.2in} 0.996
	&\hspace*{0.25in} 0.00008 &\hspace*{0.2in} $-14^\circ$\\
  \hspace*{0.15in} 1.7 &\hspace*{0.2in} 3.4 & $4.0 \times 10^{-5}
	$\hspace*{0.45in} & $3.2 \times 10^{-3}$ 
        &\hspace*{0.2in} 0.33 &\hspace*{0.2in} 0.75&\hspace*{0.2in} 0.992
	&\hspace*{0.25in} 0.0033 &\hspace*{0.2in} $-2^\circ$\\
  \hspace*{0.15in} 2.2 &\hspace*{0.2in} 3.5 & $8.8 \times 10^{-5}
	$\hspace*{0.45in} & $3.2 \times 10^{-3}$ 
        &\hspace*{0.2in} 0.24 &\hspace*{0.2in} 0.63&\hspace*{0.2in} 0.996
	&\hspace*{0.25in} 0.0008 &\hspace*{0.2in} $-4^\circ$\\
  \end{tabular} 
\end{table*}
For either the LMA or LOW version, $M_\nu$ is then obtained by the seesaw 
formula \cite{gmrsy}, $M_\nu = N^T M^{-1}_R N$.  With $M_\nu$ complex 
symmetric, both $M^\dagger_\nu M_\nu$ and $M_\nu$ itself can be diagonalized 
by the same unitary transformation, $U_\nu$, where in the latter case we find
\begin{equation}
	U^T_\nu M_\nu U_\nu = {\rm diag}(m_1,\ -m_2,\ m_3).
\label{eq:majdiag}
\end{equation}
\noindent With real light neutrino masses, $U_\nu$ can not be arbitrarily phase 
transformed and is uniquely specified up to sign changes on its column
eigenvectors \cite{ag}.  Hence $U_{MNS}$ is found by applying phase 
transformations on $U^\dagger_L U_\nu$ to bring $U^\dagger_L U_\nu$  into the 
parametric form of Eq. (\ref{eq:mixing}) whereby the $e1,\ e2,\ \mu 3$ and 
$\tau 3$ elements are real and positive, the real parts of the $\mu 2$ and 
$\tau 1$ elements are positive, while the real parts of the $\mu 1$ and 
$\tau 2$ elements are negative.  The inverse phase transformation of that 
applied on the right can then be identified with the Majorana phase matrix, 
$\Phi_M$ of Eq. (\ref{eq:mixing}).  The evolution of the predicted values 
between the GUT scale and the low scales can be safely ignored 
\cite{pok}, since $\tan \beta \simeq 5$ is moderately low and the neutrino 
mass spectrum is hierarchical with the opposite CP parities present in 
Eq. (\ref{eq:majdiag}).

We can now examine the viable region of GUT model parameter space that is
consistent with either the LMA
or LOW solar neutrino solution, and explore the 
predicted relationships among the observables $\sin^2 2\theta_{23},\ 
\sin^2 2\theta_{12},\  \sin^2 2\theta_{13},\  \delta_{CP}$,\ 
$\Delta m^2_{32}$, and $\Delta m^2_{21}$. We shall emphasize here the
simpler cases in which there are, in effect, only two additional real 
dimensionless GUT model parameters, $a$ and $b$ in the LMA version or $d$ and
$e$ in the LOW version.  In either version, the third parameter $\Lambda_R$ 
sets the scale of $\Delta m^2_{32}$.  The more general CP results obtained
for the LMA solution with the presence of a complex parameter $a$ in Eq. 
(\ref{eq:acomplex}) have been explored in detail in \cite{ag}.

\begin{figure*}[t!]
\includegraphics*[width=2.4in]{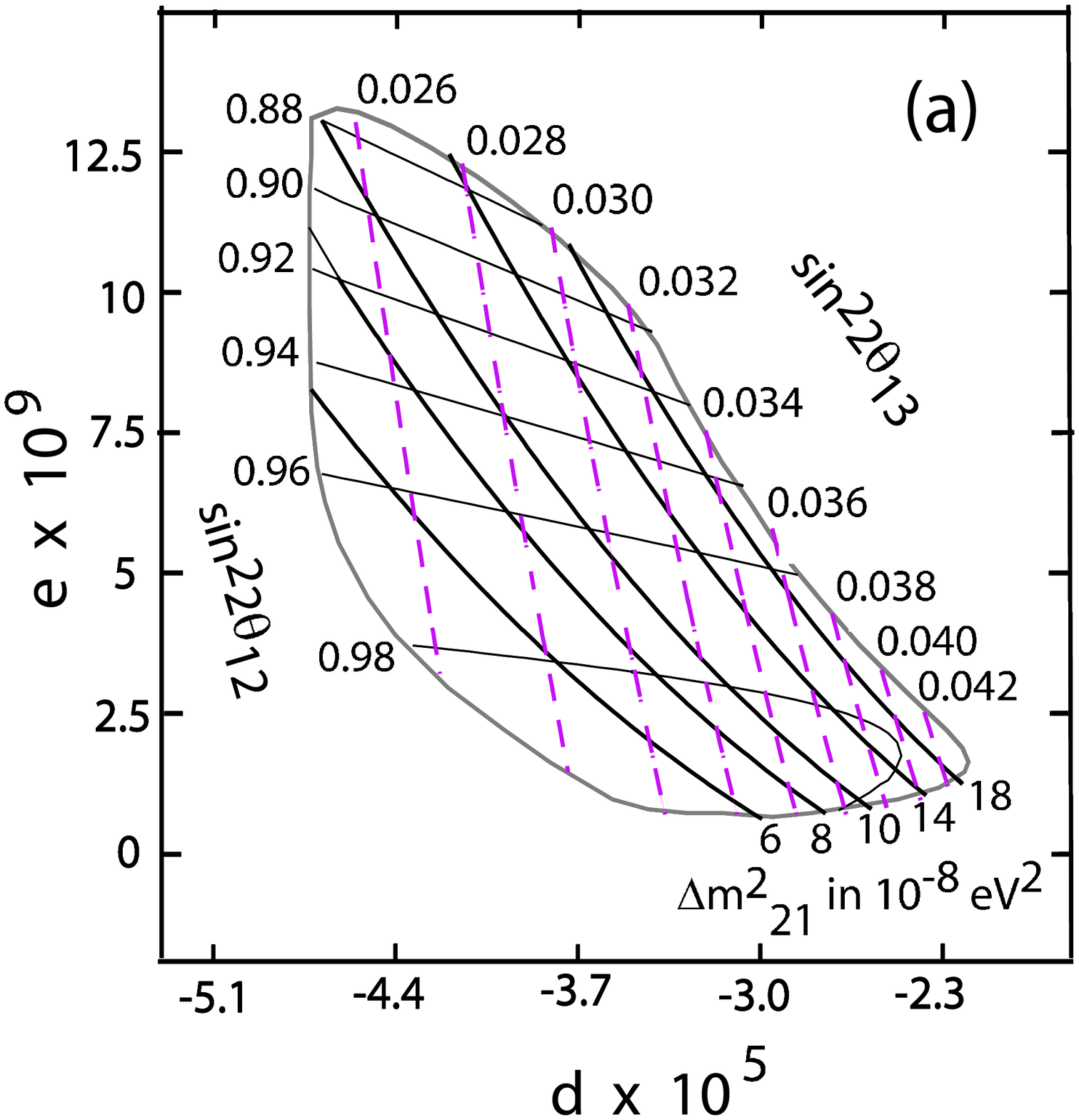}
\hspace*{0.9in}
\includegraphics*[width=2.3in]{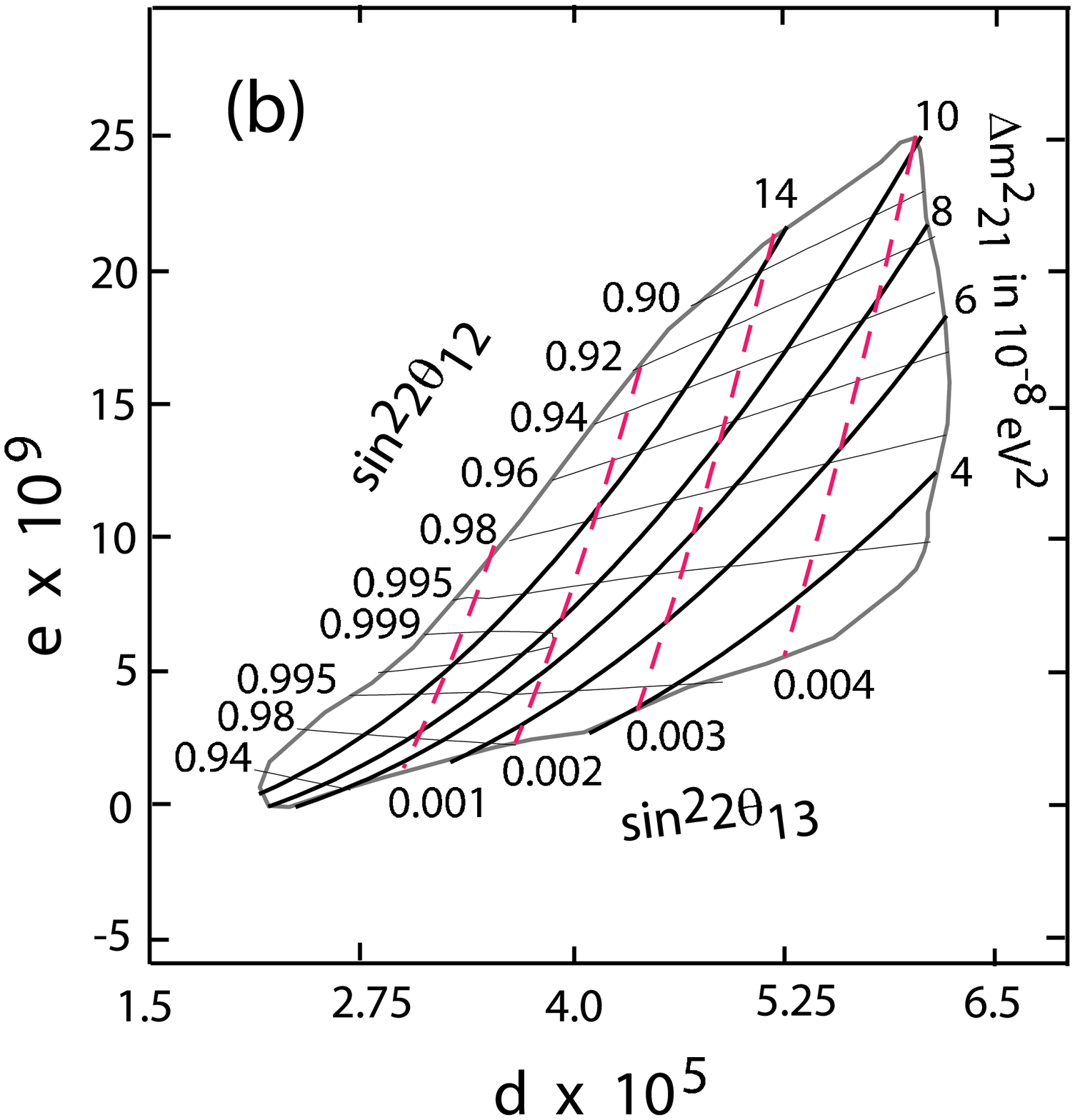}
\vspace{-0.5in}
\caption[]{The viable region of GUT parameter space consistent 
with the present bounds on the LOW MSW solution for (a) negative $d$ and 
(b) positive $d$.  Contours of constant $\sin^2 2\theta_{13}$, 
$\sin^2 2\theta_{12}$, and $\Delta m^2_{21}$ are shown.\\[-0.2in] }
\label{fig:LOW}
\end{figure*}
%
%
\begin{table*}[floatfix]
\caption[]{\label{tab:x2}
List of four points selected in the LOW allowed parameter region to illustrate
the neutrino oscillation parameter predictions of the GUT model.\\}
\begin{tabular}{rrccccccl}
   $d$\hspace*{0.3in} & $e$\hspace*{0.3in} &\hspace*{0.2in} 
	$\Delta m^2_{21}\ (eV^2)$ &\hspace*{0.2in} $\Delta m^2_{32}\ (eV^2)$ 
	&\hspace*{0.2in} 
	$\tan^2 \theta_{12}$ &\hspace*{0.2in} $\sin^2 2\theta_{12}$ 
	&\hspace*{0.2in} $\sin^2 2\theta_{23}$ &\hspace*{0.2in} 
        $\sin^2 2\theta_{13}$\\ \hline 
   $-4.2 \times 10^{-5}$ &\hspace*{0.2in} $10.0 \times 10^{-9}$ &\hspace*{0.2in}
	$1.20 \times 10^{-7}$ &\hspace*{0.2in} $3.0 \times 10^{-3}$ 
        &\hspace*{0.2in} 0.56 &\hspace*{0.2in} 0.906 
	&\hspace*{0.2in} 0.911&\hspace*{0.2in} 0.028\\
   $-4.1 \times 10^{-5}$ &\hspace*{0.2in} $4.0 \times 10^{-9}$ &\hspace*{0.2in}
	$0.52 \times 10^{-7}$ &\hspace*{0.2in} $3.0 \times 10^{-3}$ 
        &\hspace*{0.2in} 0.81 &\hspace*{0.2in} 0.975 
	&\hspace*{0.2in} 0.899&\hspace*{0.2in} 0.027\\
   $-3.6 \times 10^{-5}$ & $3.0 \times 10^{-9}$ &\hspace*{0.2in}
	$0.64 \times 10^{-7}$ &\hspace*{0.2in} $3.0 \times 10^{-3}$ 
        &\hspace*{0.2in} 0.86 &\hspace*{0.2in} 0.980 &\hspace*{0.2in} 0.898 
	&\hspace*{0.2in} 0.030 \\
   $3.6 \times 10^{-5}$ & $5.0 \times 10^{-9}$ &\hspace*{0.2in}
	$0.98\times 10^{-7}$ &\hspace*{0.2in} $3.0 \times 10^{-3}$ 
        &\hspace*{0.2in} 1.00 &\hspace*{0.2in} 0.999 &\hspace*{0.2in} 0.914 
	&\hspace*{0.2in} 0.0016\\
   $5.3 \times 10^{-5}$ & $10.0 \times 10^{-9}$ &\hspace*{0.2in}
	$0.50\times 10^{-7}$ &\hspace*{0.2in} $3.0 \times 10^{-3}$ 
        &\hspace*{0.2in} 0.82 &\hspace*{0.2in} 0.989 &\hspace*{0.2in} 0.912 
	&\hspace*{0.2in} 0.0039\\
   $5.0 \times 10^{-5}$ & $13.0 \times 10^{-9}$ &\hspace*{0.2in}
	$0.85 \times 10^{-7}$ &\hspace*{0.2in} $3.0 \times 10^{-3}$ 
        &\hspace*{0.2in} 0.70 &\hspace*{0.2in} 0.966 &\hspace*{0.2in} 0.918 
	&\hspace*{0.2in} 0.0033\\
  \end{tabular} 
\end{table*}
The viable region of GUT model parameter space consistent with the LMA solar
solution is shown in Fig.~\ref{fig:gut_plot}. Both parameters $a$ and $b$ are
constrained by the data to be close to unity, with $1.0 \aprle a \aprle 2.4$ 
and $1.8 \aprle b \aprle 5.2$. Superimposed on the allowed region,
Fig.~\ref{fig:gut_plot}(a) shows contours of constant $\sin^2 2\theta_{12}$
and contours of constant $\Delta m^2_{21}$.  
Figure \ref{fig:gut_plot}(b) similarly displays the allowed region with 
contours of constant $\sin^2 2\theta_{12}$ and $\sin^2 2\theta_{13}$ 
superimposed.  The nearly parallel nature of the contours of 
$\Delta m^2_{21}$ in (a) and $\sin^2 2\theta_{13}$ in (b) indicates a
strong correlation between them.  As the predicted $\Delta m^2_{21}$ increases,
the predicted $\sin^2 2\theta_{13}$ decreases.  Note that if the LMA solution 
is indeed the correct solution, KamLAND \cite{KamLAND} is expected 
to provide measurements of $\Delta m^2_{21}$ and $\sin^2 2\theta_{12}$ to a 
precision of about 10\% \cite{KLprec}.  From these measurements the model 
parameters $a$ and $b$ can be determined from Fig.~\ref{fig:gut_plot}(a).   
Figure~\ref{fig:gut_plot}(b) can then be used to give a prediction for 
$\sin^2 2\theta_{13}$ with a precision also of order 10\%. 

In Table~\ref{tab:x1} we have selected six points in the LMA allowed
parameter region to illustrate the neutrino oscillation 
predictions of the GUT model.  The correlations noted above are evident.  It 
is also striking how nearly maximal are the values for the atmospheric mixing 
parameter, $\sin^2 2\theta_{23}$.  This apparently arises not from some 
additionally imposed symmetry but rather from the fine tuning between the 
right-handed Majorana and Dirac neutrino mass matrices, cf. Eqs.
(\ref{eq:Dirac}) and (\ref{eq:MRLMA}).  However, if an additional phase is 
incorporated into $M_R$ for this LMA case as indicated in Eq.
(\ref{eq:acomplex}), the maximality of the atmospheric mixing is decreased 
to the lower bound in Eq. (\ref{eq:atm}) as $|\delta_{CP}|$ approaches 
$50^\circ$.  See \cite{ag} for more details.

Turning now to the GUT model version for the LOW solution, we find that 
there are two parametric regions shown in Fig. 2 for the presently allowed 
solutions, corresponding to $-4.8 \aprle d \times 10^5 \aprle -2.2,\ 
0 \aprle e \times 10^9 \aprle 13$ and $2.0 \aprle d \times 10^5 \aprle 6.0,
\ 0 \aprle e \times 10^9 \aprle 25$.  Here no dramatic correlation
between $\sin^2 2\theta_{13}$ and $\Delta m^2_{21}$ exists, so we have 
plotted contours of $\Delta m^2_{21}$, $\sin^2 2\theta_{12}$ and 
$\sin^2 2\theta_{13}$ on the same figures.  If KamLAND fails to see a signal
for the LMA region while Borexino \cite{Borex}, for example, identifies 
oscillations corresponding to the LOW region and determines 
$\sin^2 2\theta_{12}$ and $\Delta m^2_{21}$ with nearly 10\% precision 
\cite{Borprec}, 
$\sin^2 2\theta_{13}$ will be specified up to a two-fold ambiguity in the 
GUT model in question.  A first measurement of $\sin^2 2\theta_{13}$ would
resolve the ambiguity, and a precise measurement would test the model.  For 
the negative $d$ version, a SuperBeam facility capable of probing down to 
$\sin^2 2\theta_{13} \simeq 0.003$ will be able to test the model,
while for the positive $d$ version the complete parameter space can only be 
tested with a Neutrino Factory.  Table II gives the 
relevant mixing solutions for a set of six points.  In contrast to
the LMA results with small CP phases, we see that the atmospheric mixing for
the LOW solution is large but not nearly so maximal.

In conclusion, we have studied predictions for a particular but representative
GUT model that can accommodate both the LMA and LOW solar neutrino solutions. 
We find that precise measurements of $\sin^2 2\theta_{12},\ \Delta m^2_{21}$,
and $\sin^2 2\theta_{13}$ are needed to test the theory.  
Given the observed near maximal value of $\sin^2 2\theta_{23}$ the
LMA solution, which requires some fine tuning of the $M_R$ matrix, is
favored by the model. The model then predicts that the CP phase $\delta_{CP}$ 
is small and $\sin^2 2\theta_{13} \aprle 0.006$.
For the LOW solution which requires no fine tuning, $\sin^2 2\theta_{13}$ can 
be as small as this, or an order of magnitude larger, depending upon the sign 
of the $d$ model parameter in $M_R$.  Our work suggests progress on testing 
GUTs can be made with Neutrino Superbeams, but ultimately a Neutrino Factory 
will be needed to help identify the correct model.

Fermilab is operated by Universities Research Association Inc. under 
contract DE-AC02-76CH03000 with the U.S. Department of Energy. 


\end{document}